\documentclass[12pt]{article}

\begin{document}
\def\beq{\begin{eqnarray}}
\def\eeq{\end{eqnarray}}
\newcommand{\nn}{\nonumber}
\newcommand{\reals}{\mbox{${\rm I\!R }$}}
\newcommand{\nats}{\mbox{${\rm I\!N }$}}
\newcommand{\intgs}{\mbox{${\rm Z\!\!Z }$}}
\newcommand{\slambda}{\sqrt{\lambda}}
\newcommand{\ujk}{u_{j,\slambda}(1)}
\newcommand{\ujik}{u_{j,ik}(1)}
\newcommand{\uea}{u_{1,\slambda}(0)}
\newcommand{\uza}{u_{2,\slambda}(0)}
\newcommand{\ueb}{u_{1,\slambda}(1)}
\newcommand{\uzb}{u_{2,\slambda}(1)}
\newcommand{\vea}{v_{1,\slambda}(0)}
\newcommand{\vza}{v_{2,\slambda}(0)}
\newcommand{\veb}{v_{1,\slambda}(1)}
\newcommand{\vzb}{v_{2,\slambda}(1)}
\newcommand{\tk}{\frac{d}{d\lambda}}
\newcommand{\pk}{\frac{\partial}{\partial \lambda}}
\newcommand{\uja}{u_{j,\slambda} (0)}
\newcommand{\ujb}{u_{j,\slambda} (1)}
\newcommand{\vja}{v_{j,\slambda} (0)}
\newcommand{\vjb}{v_{j,\slambda} (1)}
\newcommand{\ujap}{u_{j,\slambda}' (0)}
\newcommand{\ujbp}{u_{j,\slambda}' (1)}
\newcommand{\ujx}{u_{j,\slambda} (x)}
\newcommand{\vjx}{v_{j,\slambda} (x)}
\newcommand{\ujxp}{u_{j,\slambda}' (x)}
\newcommand{\vjxp}{v_{j,\slambda}' (x)}
\newcommand{\ujxe}{u_{j,\slambda}^{(1)} (x)}
\newcommand{\ujxz}{u_{j,\slambda}^{(2)} (x)}
\newcommand{\ujxep}{u_{j,\slambda}^{(1)\prime} (x)}
\newcommand{\ujxzp}{u_{j,\slambda}^{(2)\prime} (x)}
\newcommand{\vjxe}{v_{j,\slambda}^{(1)} (x)}
\newcommand{\vjxz}{v_{j,\slambda}^{(2)} (x)}
\newcommand{\vjxep}{v_{j,\slambda}^{(1)\prime} (x)}
\newcommand{\vjxzp}{v_{j,\slambda}^{(2)\prime} (x)}
\newcommand{\drel}{\det (M+N E_{1,\slambda} (1))}
\newcommand{\sca}{\lambda \langle u_{1,0} | u_{1,\slambda} \rangle }
\newcommand{\scaneu}{\lambda \langle u_{1,0} | v_{1,\slambda} \rangle }
\newcommand{\kk}{\lambda}
\newcommand{\skk}{\sqrt{\lambda}}
\newcommand{\sla}{\sqrt{\lambda}}
\newcommand{\eop}{1_{r\times r}}
\newcommand{\nop}{0_{r\times r}}

\title{Functional determinants for general Sturm-Liouville problems}
\author{Klaus Kirsten\\
$ $\\
Department of Mathematics, Baylor University,\\
Waco TX 76798, USA\\
$ $\\
Alan J. McKane\\
$ $\\
Department of Theoretical Physics,\\
University of Manchester, Manchester M13 9PL\\
 England}
\date{\today}
\maketitle
\begin{abstract}
Simple and analytically tractable expressions for functional
determinants are known to exist for many cases of interest. We
extend the range of situations for which these hold to cover
systems of self-adjoint operators of the Sturm-Liouville type with
arbitrary linear boundary conditions. The results hold whether or
not the operators have negative eigenvalues. The physically
important case of functional determinants of operators with a zero
mode, but where that mode has been extracted, is studied in detail
for the same range of situations as when no zero mode exists. The
method of proof uses the properties of generalised zeta-functions.
The general form of the final results are the same for the entire
range of problems considered.
\end{abstract}

\newpage

\section{Introduction}
\label{intro}

This paper is concerned with the rather elegant, and surprisingly simple,
expressions that exist for the functional determinants of certain types of
differential operators. In an earlier paper~\cite{kir03}, we introduced a new
method for deriving these expressions for operators of a relatively simple
kind, which only used elementary ideas from complex analysis and the theory
of differential equations. Here we extend the class of problems which may be
analysed using this technique. Although the discussion necessarily becomes
more technical, the essential points remain the same, and we are able to
derive the desired results without the need for any very sophisticated
machinery.

The derivation of formulae of this kind is a topic which has been investigated
by numerous authors in the past. In our earlier paper~\cite{kir03}, we gave a
brief history of the subject. Essentially, most of the early results were
obtained by theoretical physicists who were typically interested in the
expressions obtained when carrying out Gaussian functional
integrals~\cite{gel60}-\cite{col85}. These results were then extended and
elaborated in a number of ways~\cite{lev77}-\cite{fal99}. However, many of
these latter treatments were quite abstract, and also did not deal with the
case where the operator has a zero eigenvalue (a ``zero mode''). This
situation is quite commonly encountered in real problems in theoretical
physics, since in many cases a continuous symmetry in the problem is broken,
and a zero mode is generated by Goldstone's theorem~\cite{raj82}. Although
there has been some work carried out to determine the form of functional
determinants with zero modes excluded~\cite{mck95}-\cite{kle99}, the methods
that were used involved the use of a regularisation procedure which could
have produced results which were not independent of the scheme adopted.

These were the motivations for our approach described in
Ref.~\cite{kir03}. The method used a generalised
zeta-function~\cite{bor96a}-\cite{kir01} to calculate the
functional determinants, but the analysis involved only methods
which are familiar to theoretical physicists. It also covered the
physically interesting situation where operators had zero modes
which were excluded from the evaluation of the functional
determinants. The method was described for simple operators of the
type $- d^{2}/dx^{2} + R(x)$, but for general linear boundary
conditions. In the present paper we extend this treatment in
several ways. Firstly, we derive the results for the general
Sturm-Liouville operator $- d/dx\left( P(x) d/dx \right) + R(x)$.
Secondly, we allow for the fact that operators will, in general,
have negative eigenvalues. Thirdly, we generalise the entire
formalism to systems of second-order operators. In all cases we
derive the results for functional determinants of operators which
do not have a zero mode, and for those which do, but where it has
been extracted.

The outline of the paper is as follows. In Section \ref{simple} we discuss
the formalism for the general Sturm-Liouville operator, modifying our
previous treatment to cover the case of arbitrary $P(x)>0$ and operators
with negative eigenvalues. We restrict ourselves to operators with no zero
mode; this case is discussed in Section \ref{zeromode}. In Section 4 it is
shown how the results of these two sections carry over to systems of $r>1$
degrees of freedom. We conclude in Section 5 with a summary of the results
of the paper and suggestions for future work. There are three appendices. In
Appendix \ref{appA} we discuss the conditions which have to be imposed so
that the operator is self-adjoint and give details of some technical
calculations that are required in the development of the theory. In
Appendix \ref{appB} some results on the asymptotic form of solutions of
differential equations, which are used in the main text, are derived. In
Appendix \ref{appC} some of the more technical aspects of dealing with
zero modes in systems of differential equations are presented.

\section{One-component}
\label{simple}

In this section we will describe our approach in the context of
operators of the form
\beq
L_j = -\frac{d}{dx} \left( P_{j} (x) \frac{d}{dx} \right) + R_j  (x) ,
\label{op1}
\eeq
on the interval $I=[0,1]$. The structure displayed in (\ref{op1}) is
the most general that is possible for a self-adjoint second order
differential operator of the Sturm-Liouville type. The functions
$P_{j} (x)$ and $R_{j} (x)$ are assumed to be continuous on the
interval $I$. In addition, we assume that the metric, $P_{j}(x)$,
is positive throughout the interval under consideration.
The index $j$ takes on only two values: the
operator $L_1$ is the real focus of interest, but in order to
control divergences in $\det L_1$, we actually consider the ratio
$\det L_{1}/\det L_{2}$, where $L_2$ is appropriately chosen.
Typically $L_2$ will be taken to be ``simple'' in a sense that it
can act as a reference with which $\det L_1$ can be compared.

The eigenproblem corresponding to (\ref{op1}) is \beq L_{j} \ujx =
\lambda \ujx . \label{eigenprob1} \eeq Note the symmetry
$u_{j,\skk} = u_{j,-\skk}$. It is convenient to go over to a first
order formalism and in order to have the most natural formulation
we define a new function $\vjx \equiv P_{j} (x) \ujxp$. Then from
(\ref{op1}) we have that \beq \frac d {dx} {\ujx \choose \vjx} =
\left(
\begin{array}{cc}
 0 & P_{j}^{-1} (x)\\
R_{j} (x) -\kk & 0
\end{array}
\right) { \ujx \choose \vjx} . \label{first_order} \eeq We will
adopt the notation \beq {\bf u}_{j,\skk}(x) = {\ujx \choose \vjx}
\ \ ; \ \ D_{j,\kk} = \left(
\begin{array}{cc}
 0 & P_{j}^{-1} (x)\\
R_{j} (x) -\kk & 0
\end{array}
\right)\,, \label{first_order_ele} \eeq in which case
(\ref{first_order}) may be written as \beq \frac{d{\bf u}_{j,\skk}
(x)}{dx} = D_{j,\kk} (x) {\bf u}_{j,\skk} (x)\,.
\label{first_order_eqn} \eeq It is useful at this stage to
introduce two unique, independent solutions of the differential
equation (\ref{eigenprob1}). The solutions are made unique by
specifying the ``initial conditions'', that is, the value of the
solutions and their derivatives at $x=0$. Denoting these two
solutions by $\ujxe$ and $\ujxz$, the most general solution of
(\ref{eigenprob1}) may then be expressed as \beq { \ujx \choose
\vjx} &=& \alpha { \ujxe \choose \vjxe } + \beta
{ \ujxz \choose \vjxz } \nonumber \\
&=&
\left(
\begin{array}{cc}
u_{j,\skk}^{(1)} (x) & \ujxz \\
\vjxe & \vjxz
\end{array}
\right)
\left(
\begin{array}{c}
\alpha \\
\beta
\end{array}
\right)\,. \label{general} \eeq We now define the two matrices
\beq E_{j,\skk} (x) = \left(
\begin{array}{cc}
u_{j,\skk}^{(1)} (x) & \ujxz \\
\vjxe & \vjxz
\end{array}
\right) \ \ ; \ \ H_{j,\skk} (x) = \left(
\begin{array}{cc}
u_{j,\skk}^{(1)} (x) & \ujxz \\
\ujxep & \ujxzp
\end{array}
\right)\,, \label{EandH} \eeq which are related by \beq E_{j,\skk}
(x) = \left(
\begin{array}{cc}
1 & 0 \\
0 & P_{j} (x)
\end{array}
\right) H_{j,\skk} (x)\,. \label{EtoH} \eeq It follows that $\det
E_{j,\skk} (x) = P_{j} (x) \det H_{j,\skk} (x)$. Since $\ujxe$ and
$\ujxz$ are independent solutions, $\det H_{j,\skk} \neq 0$, and
therefore $\det E_{j,\skk} \neq 0$ because $P_{j} (x) > 0\ \forall
x$. Furthermore, because $\det E_{j,\slambda} (x)$ is the Wronski
determinant for the differential operator (\ref{first_order_ele}),
we see
that $\det E_{j,\skk} (x)$ is independent of $x$. A convenient
choice for the set of initial conditions is $E_{j,\skk} (0) =
I_{2}$ (here, and throughout the paper, $I_{m}$ is the $m \times
m$ unit matrix). From this it follows that $\det E_{j,\skk} (x) =
1 \ \forall x \in [0,1]$. Also with this choice for the initial
conditions on $\ujxe$ and $\ujxz$, it follows from (\ref{general})
that $\alpha = \uja$ and $\beta = \vja$, that is, \beq \ujx = \uja
\ujxe + \vja \ujxz\,, \label{general_sec} \eeq or in terms of the
first order formalism, \beq { \ujx \choose \vjx} = E_{j,\skk} (x)
{ \uja \choose \vja}\,. \label{general_fir} \eeq

So far no mention has been made of the boundary conditions on
(\ref{eigenprob1}). These take the form of two conditions on the
set $\left\{ \uja, \ujap, \ujb, \ujbp \right\}$. These can be
converted into conditions on ${\bf u}_{j,\skk}$ at the boundaries,
and for the case of linear boundary conditions \beq M  {\uja
\choose \vja } + N { \ujb \choose \vjb} = {0 \choose 0} ,
\label{gbc} \eeq where $M$ and $N$ are $2\times 2$ matrices whose
entries characterise the nature of the boundary conditions. Using
(\ref{general_fir}) these boundary conditions may be written as
\beq \left[ M + N E_{j,\skk} (1) \right] {\uja \choose \vja } = {
0 \choose 0}\,, \label{alt_gbc} \eeq and so the condition on $\kk$
for eigenfunctions to exist is \beq \det \left[ M + N E_{j,\skk}
(1) \right] = 0\,. \label{eigenvalues} \eeq

The equations (\ref{gbc}) are the most general linear boundary conditions.
They fall naturally into two classes:
\begin{itemize}
\item[(i)] $\det M = 0, \det N = 0$. In this case we can show that the
matrices $M$ and $N$ may be chosen to be of the form,
\beq
M = \left(
\begin{array}{cc}
A & B \\
0 & 0
\end{array}
\right)
\ \ ; \ \
N = \left(
\begin{array}{cc}
0 & 0 \\
C & D
\end{array}
\right)\,,
\label{MN_Robin}
\eeq
that is, the boundary conditions are of the Robin type: $A \uja + B\vja = 0$
and $C \ujb + D\vjb = 0$.

First, let us prove that $\det M = 0$ and $\det N = 0$ implies that
$M {\bf u}_{j,\sqrt{\lambda}} (0)=0$ and $N {\bf u}_{j,\sqrt{\lambda}} (1)=0$.
To see this, define
${\bf u}^{\prime}_{j,\sqrt{\lambda}} (x)=Q^{-1}{\bf u}_{j,\sqrt{\lambda}} (x)$
and multiply (\ref{gbc}) by $P$, where $P$ and $Q$ are arbitrary non-singular
matrices. Then defining $M^{\prime}=PMQ$ and $N^{\prime}=PNQ$, we obtain
the same boundary conditions but in the primed system. However, since $M$ has
rank 1, we may choose $P$ and $Q$ in such a way that
$M^{\prime}={\rm diag}(1, 0)$ or ${\rm diag}(0, 1)$. Furthermore, since
$N^{\prime}$ has zero determinant it must have one of the following four forms:
\beq
A_1 &=& \left(\begin{array}{cc} a & b \\
ka & kb
\end{array}\right), \quad
A_2 = \left(\begin{array}{cc} ka & kb \\
a & b
\end{array}\right), \nn\\
A_3 &= &\left(\begin{array}{cc} a & ka \\
b & kb
\end{array}\right), \ \ \ \ \,
A_4= \left(\begin{array}{cc} ka & a\\
kb & b
\end{array}\right).
\nn
\eeq
Writing out the boundary conditions explicitly when
$M^{\prime}={\rm diag}(1, 0)$ and $N^{\prime}$ has each of these four forms,
we find that for all cases where there are two independent conditions,
$u^{\prime}_{j,\sqrt{\lambda}} (0)=0$, that is,
$M^{\prime} {\bf u}^{\prime}_{j,\sqrt{\lambda}} (0)=0$. Similarly,
if $M^{\prime}={\rm diag}(0, 1)$, we find that for all four possible forms of
$N^{\prime}$, all valid boundary conditions lead to
$v^{\prime}_{j,\sqrt{\lambda}} (0)=0$, which for this choice of $M^{\prime}$
once again gives $M^{\prime} {\bf u}^{\prime}_{j,\sqrt{\lambda}} (0)=0$.
Returning to the unprimed system this implies that
$M {\bf u}_{j,\sqrt{\lambda}} (0)=0$ and so from (\ref{gbc}),
$N {\bf u}_{j,\sqrt{\lambda}} (1)=0$, as required.

We may now use the fact that $M$ and $N$ separately must have one of the forms
$A_{1},\ldots,A_{4}$. In each case there is only one independent relation of
the Robin type. This may be written in the language of $M$ and $N$ matrices
by adopting the forms (\ref{MN_Robin}).

\item[(ii)] $\det N \neq 0$. In Appendix B we show that this implies that
$\det M \neq 0$. Then, from (\ref{gbc}), \beq {\uja \choose \vja }
= - M^{-1} N { \ujb \choose \vjb}. \nn \eeq Since $N$ is not null,
neither is $M^{-1}N$, and so either $\uja$ or $\vja$ depend on the
boundary conditions at $x=1$. Boundary conditions such as these
are called two-point boundary conditions, or non-separated
boundary conditions, in contrast to the one-point or separated
boundary conditions described by (\ref{MN_Robin}).

\end{itemize}

After this short review of the background, we are now in a
position to describe our method for obtaining the basic formula
for $\det L_{1}/\det L_{2}$. The starting point is the observation
from (\ref{eigenvalues}) that the function $\det [ M + N
E_{j,\skk} (1) ]$ has zeros at values $\kk$ which are eigenvalues
of $L_j$, as given by (\ref{eigenprob1}). An alternative statement
is that the logarithmic derivative of $\det [ M + N E_{j,\skk} (1)
]$ has a simple pole with unit residue at these values of $\kk$.
This allows us to define the zeta function of $L_j$ by \beq
\zeta_{L_{j}} (s) = \frac{1}{2\pi i}\,\int_{\gamma} d\kk\,\kk^{-s}
\frac{d\ }{d\kk} \ln \det \left[ M + N E_{j,\skk} (1) \right]\, ,
\label{zeta_def} \eeq where the contour $\gamma$ is
counterclockwise and encloses all eigenvalues as shown in Figure
1.

\begin{figure}[ht]
\setlength{\unitlength}{1cm}

\begin{center}

\begin{picture}(20,10)(0,0)

\put(0,0){\setlength{\unitlength}{1.0cm}
\begin{picture}(10,7.5)
\thicklines

\put(0,4){\vector(1,0){10}} \put(5.0,0){\vector(0,1){8}}
\put(10.0,4){\oval(9.4,1)[tl]} \put(5,4){\oval(0.6,0.5)[b]}
\put(4.4,4){\oval(0.6,1)[tr]} \put(4.4,4){\oval(3.0,1)[l]}
\put(7.5,4.5){\vector(-1,0){.4}}\put(4.4,3.5){\line(1,0){5.6}}
\put(5.0,4.0){\line(-1,2){2.0}} \put(.7,8){{\bf cut for
$\lambda^{-s}$}} \multiput(5.6,4)(.4,0){10}{\circle*{.15}}
\multiput(3.5,4)(.4,0){3}{\circle*{.15}}
\put(8.0,7.5){{\bf $\kk$-plane}}

\end{picture}}

\end{picture}

\caption{\label{fig1}Contour $\gamma$ in the complex plane.}

\end{center}

\end{figure}
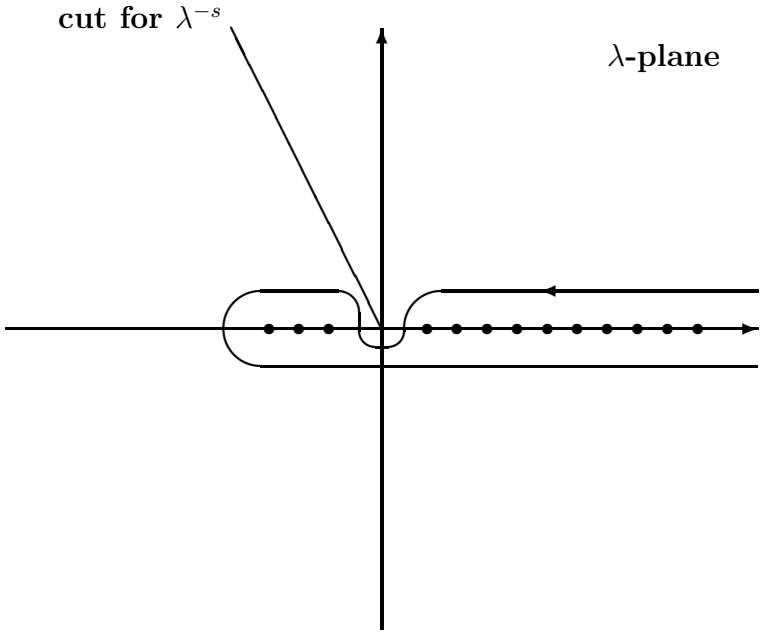

As given, the representation is valid for $\Re s > 1/2$. In this
section we assume that there are no zero modes, but note that we
allow for negative eigenvalues. In order to avoid the negative
eigenvalues lying on the cut of the complex square root, we define
the branch cut to be at an angle $\theta$ to the positive real
axis. For most applications it is the ratio of determinants of two
operators that naturally occurs. This is found by analysing \beq
\zeta_{L_{1}} (s) - \zeta_{L_{2}} (s) = \frac{1}{2\pi
i}\,\int_{\gamma} d\kk\,\kk^{-s} \frac{d\ }{d\kk}\ln \frac{\det
\left[ M + N E_{1,\skk} (1) \right]} {\det \left[ M + N E_{2,\skk}
(1) \right]}\,. \label{zeta_diff} \eeq The first idea is to deform
the contour such that it encloses the branch cut of
$\lambda^{-s}$. In order to see in which range of $s$-values this
is possible, let us consider the large-$\Im \slambda $ behaviour
of the integrand.

As shown in Appendix $B$, we have for $P_1(x) = P_2 (x)$ as
$\Im\slambda  \to \pm\infty$ the behaviour \beq \frac d
{d\lambda}\ln \frac{\det \left[ M + N E_{1,\skk} (1) \right]}
{\det \left[ M + N E_{2,\skk} (1) \right]} = {\cal O} \left( \frac
1 {\lambda^{3/2}}\right).\label{asympquot} \eeq So for $-1/2 < \Re
s < 1$ we can shift the contour such as to enclose the cut and
ultimately we can shrink it to the cut. Taking due care of the
definition of the complex root near the cut, we find for the upper
part \beq \zeta^u _{L_1} (s) - \zeta^u _{L_2} (s) &=& -\frac 1
{2\pi i} e^{-is\theta } \int\limits_0^\infty d\lambda \,\,
\lambda^{-s} \frac d {d\kk} \ln \frac{\det \left[ M + N
E_{1,e^{i\theta/2} \skk} (1) \right]} {\det \left[ M + N
E_{2,e^{i\theta/2 }\skk} (1) \right]},\nn\eeq whereas for the
lower part we have \beq \zeta^l _{L_1} (s) - \zeta ^l_{L_2} (s)
&=& \frac 1 {2\pi i} e^{is (2\pi - \theta )}\int\limits_0^\infty
d\lambda \,\, \lambda^{-s} \frac d {d\kk} \ln \frac{\det \left[ M
+ N E_{1,e^{-i(\pi - \theta /2)} \skk} (1) \right]} {\det \left[ M
+ N E_{2,e^{-i(\pi - \theta /2) }\skk} (1) \right]}.\nn\eeq Using
the symmetry $E_{j,e^{i\theta /2} \skk} (1) = E_{j,e^{-i (\pi
-\theta /2) } \skk} (1)$, these contributions add up to  yield
 \beq \zeta _{L_1} (s) - \zeta _{L_2} (s) &=& \frac 1 {2\pi i}
\left( e^{is
(2\pi - \theta )} - e^{-is\theta } \right) \times \nn\\
& &\int\limits_0^\infty d\lambda \,\, \lambda^{-s} \frac d {d\kk}
\ln \frac{\det \left[ M + N E_{1,e^{i\theta/2} \skk} (1) \right]}
{\det \left[ M + N E_{2,e^{i\theta /2}\skk} (1) \right]}\nn\\
&=& e^{is(\pi -\theta ) } \frac{\sin (\pi s )} \pi
\int\limits_0^\infty d\lambda \,\, \lambda^{-s} \frac d {d\kk} \ln
\frac{\det \left[ M + N E_{1,e^{i\theta /2} \skk} (1) \right]}
{\det \left[ M + N E_{2,e^{i\theta /2}\skk} (1)
\right]}\label{zetcont}\eeq For $\theta = \pi$ and $P_j (x) =1$
this reduces to the result of our previous paper \cite{kir03}.
This type of result is now perfectly suited for the evaluation of
the determinant quotient. The prefactor disappears at $s=0$ and so
\beq \zeta ' _{L_1} (0) - \zeta ' _{L_2} (0) = -\ln \frac{\det
\left[ M + N E_{1,0} (1) \right]} {\det \left[ M + N E_{2,0} (1)
\right]}.\label{detquot}\eeq In particular, we note that the
answer does not depend on the angle $\theta$. Simplifying notation
we define \beq y_{j} (x) = \lim_{\skk \to 0} \ujx , \ \
y_{j}^{(a)} (x) = \lim_{\skk \to 0} u_{j,\skk}^{(a)} (x) \ (a=1,2)
, \ \ Y_{j} (x) = \lim_{\skk \to 0} E_{j,\skk} (x)\,.
\label{y_defs} \eeq We will refer to $y_{j} (x)$ and $y_{j}^{(a)}
(x)$ as homogeneous solutions since they all satisfy the equation
$L_{j} y_{j} (x) =0$. Then the result is from (\ref{detquot}) \beq
\frac{\det L_{1}}{\det L_{2}} = \frac{\det \left[ M + N Y_{1} (1)
\right]}{\det \left[ M + N Y_{2} (1) \right]} .\label{result1}
\eeq This is formally identical to the result we obtained when
$P_{j} (x)=1$ and all of the eigenvalues were
positive~\cite{kir03}. This shows that this simple result is
obtained even with the added complications of non-trivial metrics
and negative eigenvalues, as long as $L_2$ is chosen so that
$P_{2} (x)=P_{1} (x)$.

\section{Determinants with zero modes extracted}
\label{zeromode}

In this section we discuss the evaluation of determinants of
operators which have a zero eigenvalue and where this eigenvalue
has been extracted in the definition of the determinant. We shall
indicate this exclusion with a prime: thus $\det ' L$ will denote
the determinant of the operator $L$ with the zero mode extracted.
Clearly the method used in the last section to derive the formula
for the ratio of determinants runs into difficulty when evaluating
such determinants. Even if the contour $\gamma$ is chosen to only
surround the non-zero values of $\lambda$, the deformation of this
contour to the branch cut will encounter the pole at the origin.
Rather than dealing directly with this extra singularity, we can
look for a function which vanishes at all the non-zero values of
$\lambda$, but not at the zero eigenvalue. This function can then
be used as the basis of the definition of a (modified) zeta-function,
from which $\det ' L$ can be calculated. As we will show in this
section, the quantity $f_{1,\skk} \equiv (-1/\lambda)\,\drel$ has
these properties: it clearly vanishes at all the required non-zero
values of $\lambda$ by (\ref{eigenvalues}) and we will show that
$f_{1,0} \neq 0$. Since we will assume that the ``normalising''
operator $L_2$ has no zero modes, $j$ will be set equal to $1$
throughout any discussion involving zero modes.

The first step in the derivation is relevant even if there is no
zero mode. It consists of demanding that the solution $\ujx =
\alpha \ujxe + \beta \ujxz$ satisfies one of the boundary
conditions. We may choose either one of the conditions to be
satisfied, but in general it will fix the functional form of
$\ujx$ by determining the ratio of $\alpha$ to $\beta$. In the
special cases of Dirichlet and Neumann boundary conditions it will
result in $\alpha$ and $\beta$, respectively, being set equal to
zero. Although the normalisation of $\ujx$ is obviously left
undetermined, we shall now show that a \emph{suitable choice of
normalisation results in a significant simplification of the
analysis}.

To see this let us write out the boundary conditions (\ref{gbc}) in full:
\beq
m_{11}\uja + m_{12}\vja + n_{11}\ujb + n_{12}\vjb &=& 0 \label{gbc_1} \\
m_{21}\uja + m_{22}\vja + n_{21}\ujb + n_{22}\vjb &=& 0\,.
\label{gbc_2} \eeq Now consider the explicit form of the matrix $M
+ N E_{j,\skk} (1)$: \beq \left(
\begin{array}{cc}
m_{11} + n_{11}u_{j,\skk}^{(1)} (1) + n_{12}v_{j,\skk}^{(1)} (1) \ &
\ m_{12} + n_{11}u_{j,\skk}^{(2)} (1) + n_{12}v_{j,\skk}^{(2)} (1) \\ \\
m_{21} + n_{21}u_{j,\skk}^{(1)} (1) + n_{22}v_{j,\skk}^{(1)} (1) \
& \ m_{22} + n_{21}u_{j,\skk}^{(2)} (1) + n_{22}v_{j,\skk}^{(2)}
(1)
\end{array}
\right)\,,
\label{M+NE}
\eeq
where we have used the definition of $E_{j,\skk}$ given in (\ref{EandH}). If
we add $\beta/\alpha$ times column 2 to column 1 of (\ref{M+NE}) we get a
second matrix whose first column is just $\alpha^{-1}$ times the boundary
conditions given in (\ref{gbc_1}) and (\ref{gbc_2}) (remembering that
$E_{j,\skk} (0) = I_{2}$). So suppose we ask that only the boundary condition
(\ref{gbc_1}) is satisfied. Since the determinants of both these matrices
are equal, it follows that
\beq
\det (M+N E_{j,\skk} (1)) &=& \alpha^{-1}\,\left\{ m_{21}\uja
+ m_{22}\vja + n_{21}\ujb + n_{22}\vjb \right\} \nonumber \\
&\times& (-1) \left( m_{12} + n_{11} u_{j,\skk}^{(2)} (1)  +
n_{12} v_{j,\skk}^{(2)} (1) \right)\,. \nonumber \eeq Therefore if
we make the choice \beq \alpha = - \left( m_{12} + n_{11}
u_{j,\skk}^{(2)} (1)  + n_{12} v_{j,\skk}^{(2)} (1) \right)\,,
\label{alpha} \eeq then \beq \det (M+N E_{j,\skk} (1)) =
m_{21}\uja + m_{22}\vja + n_{21}\ujb + n_{22}\vjb\,.
\label{modified_bc} \eeq Similarly if we add $\alpha/\beta$ of
column 1 to column 2 of (\ref{M+NE}) we get another matrix whose
second column is just $\beta^{-1}$ times the boundary conditions
given in (\ref{gbc_1}) and (\ref{gbc_2}). Again choosing the first
boundary condition to be satisfied, and also now asking that
(\ref{modified_bc}) holds, then we determine $\beta$ to be given
by \beq \beta = \left( m_{11} + n_{11} u_{j,\skk}^{(1)} (1) +
n_{12} v_{j,\skk}^{(1)} (1) \right)\,. \label{beta} \eeq

So in summary, we have shown that if we take a solution of (\ref{eigenprob1})
of the form
\beq
\ujx = &-& \left( m_{12} + n_{11} u_{j,\skk}^{(2)} (1)  +
n_{12} v_{j,\skk}^{(2)} (1) \right) \ujxe \nonumber \\ \nonumber \\
&+& \left( m_{11} + n_{11} u_{j,\skk}^{(1)} (1) +
n_{12} v_{j,\skk}^{(1)} (1) \right) \ujxz\,,
\label{normalised_soln}
\eeq
then
\beq
M  {\uja \choose \vja } + N { \ujb \choose \vjb} =
{0 \choose \det (M+N E_{j,\skk} (1))}\,.
\label{modified_gbcs}
\eeq
That is, if $\ujx$ is chosen to satisfy only one boundary condition (in this
case the first), and its normalisation is chosen appropriately, then
$\det (M+N E_{j,\skk} (1))$ will be directly proportional to the expression
on the left hand side of the boundary condition (\ref{gbc_2}), with a
constant of proportionality equal to unity. As far as we are concerned in
this paper, the relation (\ref{modified_gbcs}) has two important consequences:
\begin{itemize}
\item[(i)] If there is no zero mode in the problem, we can simply take the
limit $\lambda \to 0$ in the above formulae and get a simplified, and more
explicit, expression for the result (\ref{result1}). To do this we choose the
particular solution of the homogeneous equation $L_{j} y_{j} (x) = 0$ to be
\beq
y_{j} (x) = &-& \left( m_{12} + n_{11} y_{j}^{(2)} (1)  +
n_{12} P_{j} (1) y_{j}^{(2)\,\prime} (1) \right) y_{j}^{(1)} (x) \nonumber \\
\nonumber \\
&+& \left( m_{11} + n_{11} y_{j}^{(1)} (1) + n_{12} P_{j} (1)
y_{j}^{(1)\,\prime} (1) \right) y_{j}^{(2)} (x)\,.
\label{normalised_homo_soln} \eeq In terms of this particular
solution, the $\lambda \to 0$ limit of (\ref{modified_bc}) may be
used to write the result (\ref{result1}) as \beq \frac{\det
L_{1}}{\det L_{2}} = \frac{m_{21} y_{1} (0) + m_{22} P_1 (0)
y_{1}^{\prime} (0) + n_{21} y_{1} (1) + n_{22} P_1 (1)
y_{1}^{\prime} (1)} {m_{21} y_{2} (0) + m_{22} P_2 (0)
y_{2}^{\prime} (0) + n_{21} y_{2} (1) + n_{22} P_2 (1)
y_{2}^{\prime} (1)}\,. \label{result2} \eeq

\item[(ii)] If there is a zero mode, instead of taking the limit
$\lambda \to 0$, we use (\ref{modified_bc}) as the source of the
two relationships we need to show, namely that $\drel \sim
\lambda$ for small $|\lambda|$, and in particular that
$f_{1,\skk}$ defined earlier, satisfies $f_{1,0} \neq 0$. We now
discuss in more detail how this is carried out.
\end{itemize}

Let us begin by defining the Hilbert space product of $u_{1,\skk}
(x)$ and $u_{1,0} (x)$ on ${\cal L}^{2} (I)$ by \beq \langle
u_{1,0} | u_{1,\skk} \rangle = \int^{1}_{0} dx\, u_{1,0} (x)^{*}
u_{1,\skk} (x)\,, \label{defn_prod} \eeq where $*$ denotes complex
conjugation. So multiplying (\ref{eigenprob1}) by $u_{1,0}
(x)^{*}$ and integrating gives \beq \int^{1}_{0} dx\, u_{1,0}
(x)^{*} L_{1} u_{1,\skk} (x) = \lambda\,\langle u_{1,0} |
u_{1,\skk} \rangle\,. \nonumber \eeq
By partial integration we get boundary terms plus $L_1 u_{1,0} (x)
^*$. This latter term is zero, so therefore \beq \left[ u_{1,\skk}
(x) v_{1,0} (x)^{*} - u_{1,0} (x)^{*} v_{1,\skk} (x)
\right]^{1}_{0} = \lambda\,\langle u_{1,0} | u_{1,\skk} \rangle\,.
\label{starting_point} \eeq We can now use (\ref{modified_gbcs})
to solve for two members of the set $\left\{ \uea , \vea , \ueb
,\right.$ $\left. \veb \right\}$ in terms of the other two and
$\drel$. An exactly analogous procedure is carried out on the set
$\left\{ u_{1,0} (0)^*, v_{1,0} (0)^*, u_{1,0} (1)^*, v_{1,0}
(1)^* \right\}$, but in this case $u_{1,0} (x)$ satisfies both of
the boundary conditions, and so (\ref{gbc}), rather than
(\ref{modified_gbcs}), should be used. This procedure is discussed
in more detail in Appendix A, where it is shown that substituting
the expressions for these four quantities into the left-hand side
of eqn. (\ref{starting_point}) shows that it is directly
proportional to $\drel$. The constant of proportionality (denoted
by ${\cal B}^{-1}$) is independent of $\lambda$, and only depends
on the nature of the boundary conditions and on the $\lambda=0$
solution ${\bf u}_{1,0} (x)$ at the boundaries. Therefore we may
write \beq \drel = {\cal B}\,\left[ u_{1,\skk} (x) v_{1,0} (x)^{*}
- u_{1,0} (x)^{*} v_{1,\skk} (x) \right]^{1}_{0} = {\cal
B}\,\lambda\,\langle u_{1,0} | u_{1,\skk} \rangle\,. \label{three}
\eeq It should be stressed that while $u_{1,0} (x)$ satisfies both
boundary conditions, $u_{1,\skk} (x)\ (\lambda \neq 0)$ satisfies
only one (together with a normalisation condition), in other
words, it has the form (\ref{normalised_soln}). If the other
boundary condition is imposed, $\lambda$ is restricted to take on
values for which $\lambda$ is an eigenvalue, and we see from
(\ref{three}) that the orthogonality condition $\langle u_{1,0} |
u_{1,\skk} \rangle = 0$ holds by virtue of (\ref{eigenvalues}).

The constant ${\cal B}$ is determined in Appendix A, where the related
question of the conditions for the operator to be self-adjoint, is also
discussed. The conclusions are:
\begin{itemize}
\item[(i)] If the boundary conditions are separated, so that
$\det M = \det N = 0$, then if the operator is self-adjoint, $M$ and $N$ can
always be chosen to be of the form (\ref{MN_Robin}), with $M$ and $N$ real. In
this case
\beq
{\cal B} = \frac{n_{21}}{v_{1,0} (1)^{*}}\ , \ {\rm if\ } n_{21} \neq 0\ \ ;
\ \ {\cal B} = - \frac{n_{22}}{u_{1,0} (1)^{*}}\ , \ {\rm if\ } n_{22}
\neq 0\,.
\label{B_i}
\eeq

\item[(ii)] If the boundary conditions are non-separated, so that
$\det M \neq 0, \det N \neq 0$, then if the operator is
self-adjoint, $M$ and $N$ can always be chosen so that one of them
is real, say $N=N_{R}$, and the other one a real matrix times a
phase: $M=M_{R} e^{i\alpha}, 0 \leq \alpha < 2\pi$. It also
follows that $\det M_{R} = \det N_{R}$. In this case \beq {\cal B}
= \frac{n_{12} n_{21} - n_{11} n_{22}}{n_{11} u_{1,0} (1)^{*} +
n_{12} v_{1,0} (1)^{*}}\,. \label{B_ii} \eeq
\end{itemize}

The function $f_{1,\skk}$ mentioned earlier can now be identified.
If we define \beq f_{1,\skk} \equiv - \frac{\drel}{\lambda} = -
{\cal B}\,\langle u_{1,0} | u_{1,\skk} \rangle\,, \label{eff} \eeq
we see that it vanishes at the required values of $\lambda$, but
is non-zero when $\lambda=0$. However, in our evaluation of the
contour integral in the last section, it was also vital that the
large $|\slambda|$ behaviour for $j=1$ and $j=2$ were the same, so
actually we need to replace $\drel$ in the integrand of the
contour integral by $(1-\lambda)f_{1,\slambda}$. This has the
required properties when both $\lambda =0$ and $\lambda \neq 0$, but
in addition it behaves like $\drel$ for large $|\slambda|$, also
as required. So in order to derive an expression for $\det '
L_{1}/\det L_{2}$ we need to begin from \beq \zeta_{L_{1}} (s) -
\zeta_{L_{2}} (s) = -1+\frac{1}{2\pi i}\,\int_{\gamma}
d\lambda\,\lambda^{-s} \frac{d\ }{d\lambda}\ln \frac{(1-\lambda)
f_{1,\skk}} {\det \left[ M + N E_{2,\skk } (1) \right]}\,,
\label{zeta_zm} \eeq where the contour $\gamma$ encloses the point
$\lambda = 1$ and the values of $\lambda$ on the real axis which
define the eigenvalues.

It is understood that the zero mode has been omitted from the
definition of the zeta function. For definiteness we have assumed
that the contour encloses $\lambda =1$ such that the term '-1' on
the right hand side corrects for the contribution due to the
factor $(1-\lambda )$. Proceeding as before, now noting that
$f_{1, e^{i\theta /2} \slambda }= f_{1, e^{-i (\pi -\theta /2)}
\slambda}$, we obtain \beq \zeta _{L_1} (s) - \zeta _{L_2} (s) =
-1 + e^{is (\pi -\theta )} \frac{\sin (\pi s) } \pi
\int\limits_0^\infty d\lambda \,\, \lambda^{-s} \frac d {d\lambda}
\ln \frac{ (1+e^{i\theta} \lambda) f_{1, e^{i\theta /2} \skk} } {
\det (M+N E_{2,e^{i\theta /2} \skk} (1))} .\nn\eeq For the
derivative at $s=0$ this means \beq \zeta_{L_1}' (0) - \zeta _{L_2
} ' (0) = - \ln \frac{ f_{1,0} } { \det (M+N E_{2,0} (1)) }
.\nn\eeq Using the notation of equations (\ref{y_defs}) and
(\ref{eff}), this may be cast into the final form \beq
\frac{\det ' L_{1}}{\det L_{2}} = - \frac{{\cal B} \langle y_{1} |
y_{1} \rangle} {\det \left[ M + N Y_{2} (1) \right]}.
\label{result3} \eeq

\section{Systems of differential operators}
\label{systems}

The extension from a single differential operator of the form (\ref{op1})
to a system of differential equations is relatively straightforward, the
main problem being one of notation. Provided that the previous sections have
been read, the discussion in this section should be clear, since it parallels
the case of a single operator. As an additional aid to understanding, we
illustrate new concepts which are introduced on a specific example. Some of
the more cumbersome formulae which are not vital to an overall understanding
of the formalism are relegated to Appendix \ref{appC}.

We consider the system of differential operators \beq L_j = -\frac
d {dx} \left( P_j (x) \frac d{dx} \right) I_r + R_j (x) \nn\eeq
where $R_j (x)$ is a Hermitian $r\times r$ matrix and $j=1,2$
labels the two different determinants. We assume $P_j (x)$ to be
scalar, which is the relevant case for most applications. The
second order problem is rewritten as a first order problem in the
standard way, \beq \frac d {dx} { u_{j,\sla} (x) \choose
v_{j,\sla} (x) } = D_{j,\lambda }{ u_{j,\sla} (x) \choose
v_{j,\sla} (x) }\nn\eeq with the matrix \beq D_{j,\lambda} (x) =
\left(\begin{array}{cc}
0_{r\times r} & P_j^{-1} (x) \cdot I_r \\
 R_j -\lambda \cdot I_r & \nop
 \end{array} \right),\nn\eeq
and where now $u_{j,\sla} (x)$ and $v_{j,\sla} (x)$ are
$r$-dimensional vectors. We define, as before, the fundamental
matrix as \beq E_{j,\sla} (x) = \left(
\begin{array}{ccc} u_{j,\sla}^{(1)} (x) & ... &
u_{j,\sla} ^{(2r)} (x) \\
v_{j,\sla} ^{(1)} (x) & ... & v_{j,\sla} ^{(2r)}(x) \end{array}
\right),
\nn
\eeq
with $u_{j,\sla}^{(\sigma)} (x)$, $v_{j,\sla }^{(\sigma)} (x)$,
$\sigma = 1,\ldots,2r$, being again $r$-dimensional vectors.
The boundary conditions read
\beq
M {u_{j,\sla} (0) \choose v_{j,\sla} (0) } + N
{u_{j,\sla} (1) \choose v_{ j,\sla} (1) } = {0\choose 0}
,\label{1} \eeq or, alternatively \beq (M + N E_{j,\sla} (1))
{u_{j,\sla} (0) \choose v_{j,\sla} (0) } = {0\choose
0}.\label{2}
\eeq
So the condition for the eigenvalues reads
\beq
\det ( M + N E_{j,\sla} (1) ) = 0.\label{3}
\eeq
In the case that $P_j (x)$ is scalar, the analysis in Appendix B goes
through. The only
change is that the heat kernel coefficients contain a trace over
the internal degrees of freedom. With this change, the asymptotic
behaviour of the relevant integrand is known and for $P_1 (x) =
P_2 (x)$ we can proceed as previously. In the absence of zero
modes we find formally the same answer as before, \beq \frac {\det
L_1} {\det L_2} = \frac{ \det (M+N Y_1 (1))}{\det (M + N Y_2
(1))}.\nn\eeq

Let us next consider the case with zero modes. In order to explain
the individual steps of the general formalism that we are
developing, we will illustrate each step using a specific example
encountered in the study of transition rates between metastable
states in superconducting rings \cite{tar94,tar95}. The
differential operator in this problem is defined on the interval
$[-l/2, l/2]$ and has the form \beq L_{1} = \left(
\begin{array}{cc}
- \frac{d^{2}}{dx^{2}} + (1-2\mu^{2}) & (1-\mu^{2})e^{2i\mu x} \\ \\
(1-\mu^{2})e^{-2i\mu x} & - \frac{d^{2}}{dx^{2}} + (1-2\mu^{2})
\end{array}
\right) \equiv - \,\frac{d^2}{dx^2} I_2+ R_{1} (x). \label{start2}
\eeq Boundary conditions imposed are so-called twisted boundary
conditions defined through \beq M = -\mbox{diag} ( e^{i\mu l} ,
e^{-i\mu l}, e^{i\mu l}, e^{-i \mu l} ) , \quad N = I_4.\nn \eeq
We will refer back to this example at suitable stages of our
procedure.

The starting point for the general formalism is as before, namely
(\ref{starting_point}). If we can derive a relationship of the form
\beq
\det (M+N E_{1,\sla} (1)) = {\cal B}[u_{1,\sla} (x) v_{1,0} (x) ^* - u_{1,0}
(x)^* v_{1,\sla} (x) ]^1_0\,,
\label{bit_of_three}
\eeq
where ${\cal B}$ is known, then from (\ref{starting_point}) we have that
\beq
\det (M+N E_{1,\sla} (1)) = {\cal B} \lambda\,\langle u_{1,0} | u_{1,\skk}
\rangle\,.
\label{rest_of_three}
\eeq
This is precisely as in Section \ref{zeromode}, and allows us to identify
the function $f_{1,\skk}$, defined by (\ref{eff}), which is to be used in
the proof of the result.

So, let us return to the proof of (\ref{bit_of_three}). We will show that, if
we appropriately normalise $u_{j,\sla} (x)$, then by imposing all of the
boundary conditions but one --- so that $\lambda$ is not constrained to be
an eigenvalue --- we may write
\beq
M \left( \begin{array}{cc} u_{1,\sla} (0) \\
v_{1,\sla} (0) \end{array} \right) + N\left( \begin{array}{cc} u_{1,\sla} (1)
\\
v_{1,\sla} (1) \end{array} \right) = \left( \begin{array}{c}
 0 \\ 0 \\ ... \\ 0 \\ \det \left(M+NE_{1,\sla} (1)
 \right)
\end{array}
\right) .
\label{system_bcs}
\eeq
This equation is exactly analogous to (\ref{modified_gbcs}) in Section
\ref{zeromode}, where we imposed only one out of the two boundary conditions.
Here there are $2r$ boundary conditions and we will impose $2r-1$ of them.

To obtain (\ref{system_bcs}) we first write $u_{j,\sla} (x)$ as a linear
combination of the $2r$ fundamental solutions $u^{(\sigma)}_{j,\sla} (x)$:
\beq
u_{j,\sla} (x) = \sum^{2r}_{\sigma = 1} \alpha^{(\sigma)}
u^{(\sigma)}_{j,\sla} (x) \ \ \ \Rightarrow \ \ \
v_{j,\sla} (x) = \sum^{2r}_{\sigma = 1} \alpha^{(\sigma)}
v^{(\sigma)}_{j,\sla} (x)\,,
\label{lin_comb}
\eeq
where we have dropped the $j$ and $\sla$ dependence from the $\alpha$. Since
$E_{j,\sla} (0) = I_{2r}$,
\beq
\alpha^{(\sigma)} = \left\{ \begin{array}{ll}
u_{j,\sla,\sigma} (0), & \mbox{\ if $\sigma=1,\ldots,r$} \\ \\
v_{j,\sla,\sigma - r} (0), & \mbox{\ if $\sigma=r+1,\ldots,2r$}\,,
\end{array} \right.
\nn
\eeq
where $u_{j,\sla,\sigma} (x)$ is the $\sigma$th entry of the vector
$u_{j,\sla} (x)$, with a similar notation for $v_{j,\sla} (x)$. Therefore
using (\ref{2}), but only imposing the first $2r-1$ boundary conditions, gives
\beq
( M + N E_{j,\sla} (1) )
\left( \begin{array}{c}
\alpha^{(1)} \\ \alpha^{(2)} \\ ... \\ \alpha^{(2r-1)} \\ \alpha^{(2r)}
\end{array}
\right) =
\left( \begin{array}{c}
0 \\ 0 \\ ... \\ 0 \\ \ast
\end{array}
\right)\,.
\label{eqnforcoeffs}
\eeq
First, suppose that $\det ( M + N E_{j,\sla} (1) ) \neq 0$. Then, multiplying
(\ref{eqnforcoeffs}) by $( M + N E_{j,\sla} (1) )^{-1}$ yields
\beq
\left( \begin{array}{c}
\alpha^{(1)} \\ \alpha^{(2)} \\ ... \\ \alpha^{(2r-1)} \\ \alpha^{(2r)}
\end{array}
\right) =
\frac{{\rm adj} ( M + N E_{j,\sla} (1) )}{\det ( M + N E_{j,\sla} (1) )}
\left( \begin{array}{c}
0 \\ 0 \\ ... \\ 0 \\ \ast
\end{array}
\right)\,, \nn \eeq where ${\rm adj} ( M + N E_{j,\sla} (1) )$ is
the adjoint of the matrix $M + N E_{j,\sla} (1)$. We see that the
choice of $\det ( M + N E_{j,\sla} (1) )$ for $\ast$ is natural,
since in this case the expansion coefficients have the simple form
\beq \alpha^{(\sigma)} = {\rm adj} ( M + N E_{j,\sla} (1)
)_{\sigma\,2r}\,. \label{alphas} \eeq In the $r=1$ case this
simply leads to the results (\ref{alpha}) and (\ref{beta}). If
$\det ( M + N E_{j,\sla} (1) ) = 0$, then by (\ref{3}) $u_{j,\sla}
(x)$ is an eigenfunction which satisfies the boundary conditions,
and so (\ref{system_bcs}) also holds.

Altogether there are $4r$ boundary data, $r$ data coming from each of
$u_{1,\sla} (0)$, $v_{1,\sla} (0)$, $u_{1,\sla} (1)$ and $v_{1,\sla} (1)$.
Eq. (\ref{system_bcs}) allows us to express $2r$ of the boundary data in
terms of the other $2r$ data, which we call the complementary ones. Suppose
that $b$ is a vector consisting of the $2r$ boundary data that we wish to
express by the complementary ones, collected in $b_c$. Expressing
(\ref{system_bcs}) in terms of these values gives
\beq
{\cal Z} b + {\cal Z} _c b_c =\left(
\begin{array}{c}
0 \\ 0 \\ .. \\ 0 \\ \det \left(M+NE_{1,\sla} (1)
\right)
\end{array}
\right) ,
\label{datasplit}
\eeq
where ${\cal Z}$ and ${\cal Z}_c$ are $(2r \times 2r)$ matrices built from
the various components of $M$ and $N$. To state $b$, $b_c$, ${\cal Z}$ and
${\cal Z}_c$ explicitly, we need to introduce several indices
which refer to the ways in which the boundary data are re-distributed within
each of the four boundary data groups. Let $i,j,k,l$ be indices all of which
can take on values from $0$ to $r$ and such that $i+j+k+l=2r$. Let
$\{a_1,...,a_r\}$ and $\{c_1,...,c_r\}$ be arbitrary permutations of the
numbers $\{1,...,r\}$, and also let $\{b_1,...,b_r\}$ and $\{d_1,...,d_r\}$ be
permutations of the numbers $\{r+1,...,2r\}$. These index groups are such that
$m_{a_i}$ acts on boundary data in $u_{1,\sla} (0)$, $m_{b_j}$ acts in
$v_{1,\sla} (0)$, $n_{c_k}$ acts in $u_{1,\sla} (1)$, and finally
$n_{d_l}$ acts in $v_{1,\sla} (1)$. The general form of $b, b_{c}, {\cal Z}$
and ${\cal Z}_{c}$ are discussed in Appendix~\ref{appC}, from which it is
clear that $b$ can be expressed through $b_c$ only if the matrix ${\cal Z}$
is invertible. The choice of the data $b$ has to guarantee this is indeed
the case. That this is always possible follows from the fact that $M$ and
$N$ define boundary conditions such that (\ref{system_bcs}) has a unique
solution for $\lambda$ an eigenvalue. If a suitable choice of $b$ were not
possible, the boundary value problem would not have a unique solution.

For the example described by (\ref{start2}), the most natural
choice for $b$, $b_c$, is
\beq
b = \left( \begin{array}{c}
u_{1,\sla , 1 } (l/2) \\ u_{1,\sla , 2 } (l/2) \\
v_{1,\sla , 1 } (l/2) \\ v_{1 , \sla , 2 } (l/2)  \end{array} \right),
\quad b_c = \left( \begin{array}{c}
u_{1,\sla , 1 } (-l/2) \\ u_{1,\sla , 2 } (-l/2) \\
v_{1,\sla , 1 } (-l/2) \\ v_{1 , \sla , 2 } (-l/2)  \end{array}
\right), \label{choice1} \eeq so that ${\cal Z} = N$ ($= I _4$)
and ${\cal Z} _c = M$ ($= -\mbox{diag} (e^{i\mu l}, e^{-i\mu l},
e^{i\mu l}, e^{-i \mu l} ))$. This guarantees ${\cal Z}$ is
invertible and $b$ can be expressed through $b_c$.

Alternatively one could, for instance, choose
\beq
b^{({\rm alt})} = \left(
\begin{array}{c}
u_{1,\sla , 1 } (-l/2) \\ u_{1,\sla , 2 } (-l/2) \\
v_{1,\sla , 1 } (-l/2) \\ v_{1 , \sla , 2 } (-l/2)  \end{array}
\right),\quad b_c^{({\rm alt})} = \left( \begin{array}{c}
u_{1,\sla , 1 } (l/2) \\ u_{1,\sla , 2 } (l/2) \\
v_{1,\sla , 1 } (l/2) \\ v_{1 , \sla , 2 } (l/2)  \end{array}
\right).
\label{choice2}
\eeq
In this case
$${\cal Z}^{({\rm alt})} =M, \quad {\cal Z}_c^{({\rm alt})} = N.$$
Again, ${\cal Z}$ is invertible and $b$ can be expressed through $b_c$.
Clearly, there are many other choices of $b$, $b_c$ and the
associated ${\cal Z}$, ${\cal Z}_c$.

Going back to the general formalism, given a suitable particular choice of
${\cal Z}$, this allows us to express the $2r$ data $b$ by the complementary
$2r$ data $b_c$. The explicit expression is given by equation (\ref{4}) in
Appendix~\ref{appC}. The entries of $b$ can now be substituted into the
left-hand side of (\ref{starting_point}) and the terms collected together.
As discussed in Appendix \ref{appC} this leads to (\ref{bit_of_three}) with
\beq
{\cal B}^{-1} &=& \sum_{\alpha =1} ^k {\cal
Z}_{(i+j+\alpha)(2r)} ^{-1} v_{1,0,c_\alpha} (1) ^* - \sum_{\alpha
=1} ^l{\cal
Z}_{(i+j+k+\alpha)(2r)} ^{-1} u_{1,0,d_\alpha -r} (1) ^* \nn\\
&-& \sum_{\alpha =1} ^i {\cal Z}_{\alpha (2r)}
^{-1} v_{1,0,a_\alpha} (0) ^* + \sum_{\alpha =1} ^j {\cal
Z}_{(i+\alpha) (2r)} ^{-1} u_{1,0,b_\alpha -r} (0) ^* \,,
\label{Bminus1}
\eeq
where ${\cal Z}_{\beta \gamma} ^{-1}$ refers to the $(\beta \gamma)$-component
of ${\cal Z}^{-1}$.

To illustrate the use of this result let us apply it again to the
example (\ref{start2}).

\noindent For the choice (\ref{choice1}) we have $i=0, j=0, k=2, l=2$ and we
obtain
\beq
[u_{1,\sla} (x) v_{1,0} (x) ^* - u_{1,0} (x)^*
v_{1,\sla} (x) ]^{l/2}_{-l/2} = -\det (M+N E_{1,\sla} (l/2)) u_{1,0,2} (l/2)
^*.
\nn
\eeq
For the choice (\ref{choice2}) we have $i=2,j=2,k=0,l=0$ and we obtain
\beq
[u_{1,\sla} (x) v_{1,0} (x)^* - u_{1,0} (x)^* v_{1,\sla} (x) ]^{l/2}_{-l/2} =
-\det (M+N E_{1,\sla} (l/2)) e^{i\mu l} u_{1,0,2} (-l/2) ^*.
\nn
\eeq
Comparing with (\ref{bit_of_three}) we see that for the choice (\ref{choice1})
${\cal B}^{-1} = - u_{1,0,2} (l/2) ^*$ and for the choice (\ref{choice2})
${\cal B}^{-1} = - e^{i\mu l} u_{1,0,2} (-l/2) ^*$. Taking into account the
boundary conditions for the zero mode $u_{1,0} (x)$, the two answers are seen
to agree. Furthermore, this answer agrees with the result calculated
in~\cite{tar95}.

Returning to the general case, we see that we have proved the result
(\ref{rest_of_three}) with ${\cal B}$ given by (\ref{Bminus1}). The proof
now proceeds as in Section \ref{zeromode} and we once again find the result
(\ref{result3}). The function $y_{1} (x)$ in this result is the zero mode,
and satisfies the boundary conditions, but it has to be appropriately
normalised:
\beq
y_{1} (x) = \sum^{2r}_{\sigma = 1} {\rm adj}
( M + N E_{1,0} (1) )_{\sigma\,2r}\,y_{1}^{(\sigma)} (x)\,,
\label{explicity1}
\eeq
where the $y_{1}^{(\sigma)} (x)$ are the $2r$ fundamental solutions chosen
to satisfy $Y_{1} (0) = I_{2r}$.

\section{Conclusion}
\label{conclusion}

The two main results of this paper are (\ref{result1}) and (\ref{result3}).
They give expressions for the ratio of functional determinants in terms of the
nature of the boundary conditions and the solution of the homogeneous
equations formed from the operators in question. The first result holds if
the equations have no zero modes and the second holds if such an eigenvalue
exists, but has been excluded from the evaluation of the functional
determinant. These results agree with those obtained in a previous
paper~\cite{kir03}, but now the range of operators for which they are valid
have been considerably extended to: those which have a metric
$P_{j} (x) \neq 1$, those with negative eigenvalues and systems of operators.
Although the results are simple to state, a slightly more thorough appreciation
of the method is required in order to apply them to a particular case. For
instance the solution $y_{1} (x)$, which appears in the results, is the
solution of the homogeneous equation satisfying the boundary conditions. This
solution is only defined up to a constant, but a particular choice for this
constant has to be made if the simpler form of (\ref{result1}) --- given by
(\ref{result2}) --- or the zero-mode result (\ref{result3}), is to be used.
An explicit form for $y_{1} (x)$ is given by (\ref{explicity1}). The choice
of normalisation originates from requiring that the right-hand side of
(\ref{modified_gbcs}) or (\ref{system_bcs}) is the required determinant, but
with a constant of proportionality which is equal to 1.

Once the suitably normalised solution $y_{1} (x)$ has been obtained, the rest
of the calculation is straightforward. The result only depends on this
function --- and on none of the other eigenfunctions --- and on the matrices
$M$ and $N$ which define the boundary conditions of the problem under
consideration. In addition, in some applications, the norm
$\langle y_{1} | y_{1} \rangle$ will cancel out with the Jacobian of the
transformation to collective coordinates, and therefore will not be required.
In this case, however, it will be necessary to check that the zero-mode has
the same normalisation as has been adopted in the derivation of
(\ref{result3}). So for this case only the properties of $y_{1} (x)$ at the
boundaries would be required. If an analytic expression for $y_{1} (x)$ cannot
be obtained, there should be little difficulty in obtaining numerical values
for the boundary data on this function. While the proof of the results which
we have obtained are easily accessible, they need not be understood in order
to apply the results to a particular problem.

We believe that the results presented here cover a wide range of problems
where they are likely to prove useful. There are still a number of possible
extensions that are open to investigation. Examples include operators with
derivatives higher than the second, single determinants rather than
ratios of determinants, and operators in more than one dimension. We hope
that, in addition to the concrete results which we have obtained, this paper
will serve to stimulate work on these and related problems.

\vspace{0.9cm}

{\bf Acknowledgements}: The research of K. Kirsten was partially supported
by the Max Planck Institute for Mathematics in the Sciences (Leipzig, Germany)
and the Baylor University Summer Sabbatical Program.

\newpage

\renewcommand{\theequation}{\Alph{section}\arabic{equation}}
\setcounter{section}{0} \setcounter{equation}{0}

\begin{appendix}

\section{Self-adjoint condition and related questions}
\label{appA}

In this appendix we will consider two technical points encountered
in Section \ref{zeromode}. They are
\begin{itemize}
\item[1.] {\em Condition for problem to be self-adjoint}.

The boundary conditions considered in section \ref{zeromode} are given by
(\ref{gbc_1}) and (\ref{gbc_2}):
\beq
m_{11}\uja + m_{12}\vja + n_{11}\ujb + n_{12}\vjb &=& 0 \nn \\
m_{21}\uja + m_{22}\vja + n_{21}\ujb + n_{22}\vjb &=& 0\,.
\label{gbc_appA}
\eeq
Suppose that $U_{j,\skk}^{\rm (I)} (x)$ and $U_{j,\skk}^{\rm (II)} (x)$ are
any two functions (which are not, in general, solutions of (\ref{eigenprob1}))
which satisfy these boundary conditions. The condition for the problem to be
self-adjoint is that
\beq
\left[ U_{j,\skk}^{\rm (I)} (x) V_{j,\skk}^{\rm (II)} (x)^{*} -
U_{j,\skk}^{\rm (II)} (x)^{*} V_{j,\skk}^{\rm (I)} (x) \right]^{1}_{0}
= 0\,,
\label{SAC}
\eeq
where, as in the main text,
$V_{j,\skk} (x) = P_{j} (x) U_{j,\skk}^{\prime} (x)$. We wish to solve for
any two members of the set
$\left\{ U_{j,\skk}^{\rm (I)} (0), V_{j,\skk}^{\rm (I)} (0),
U_{j,\skk}^{\rm (I)} (1), V_{j,\skk}^{\rm (I)} (1) \right\}$ in terms of
the other two by using the two boundary conditions (and similarly for the
second solution II). Substituting these four functions into (\ref{SAC}) in
terms of the other four will give us the conditions that need to be imposed
on the matrices $M$ and $N$ for the problem to be self-adjoint.

\item[2.] {\em Proof of the first equality in eqn. (\ref{three}).}

In this case the $\lambda=0$ solution $u_{1,0} (x)$ will satisfy
the boundary conditions (\ref{gbc_appA}), but the $\lambda \neq 0$
solution will only satisfy one boundary condition and a
normalisation condition, that is (see eqn. (\ref{modified_gbcs})):
\beq m_{11} \uea + m_{12} \vea + n_{11} \ueb + n_{12} \veb &=& 0
\label{modified_appA} \\
m_{21} \uea + m_{22} \vea + n_{21} \ueb + n_{22}  \veb &=& \drel
\nn \eeq As discussed in Section \ref{zeromode}, we wish to solve
for any two members of the set $\left\{ \uea , \vea , \ueb , \veb
\right\}$ in terms of the other two and $\drel$ by using the
conditions (\ref{modified_appA}). Similarly, we wish to solve for
any two members of the set $\left\{ u_{1,0} (0)^*, v_{1,0} (0)^*,
u_{1,0} (1)^*, v_{1,0} (1)^* \right\}$, in terms of the other two,
this time using (\ref{gbc_appA}). Substituting these four
functions into the left-hand side of (\ref{starting_point}), in
terms of the other four, will enable us to show that \beq \left[
u_{1,\skk} (x) v_{1,0} (x)^{*} - u_{1,0} (x)^{*} v_{1,\skk} (x)
\right]^{1}_{0} \propto \drel\,, \label{prop_appA} \eeq where the
constant of proportionality is independent of $\lambda$, and only
depends on the nature of the boundary conditions and on the
$\lambda=0$ solution ${\bf u}_{1,0} (x)$ at the boundaries.
\end{itemize}
It is clear that these two questions are related. In fact, the
proof of the first point is a special case of the proof of the
second; we simply need to set $\drel$ equal to zero everywhere. We
will prove the result first in the case of separated boundary
conditions and then for non-separated ones.

\begin{itemize}

\item[(i)] For separated boundary conditions we have $\det N = 0$
and $\det M =0$, and $M$ and $N$ may be chosen to have the form
(\ref{MN_Robin}).

\medskip

If $m_{12} \neq 0$ and $n_{22} \neq 0$, then (\ref{modified_appA}) may be
written in the form
\beq
\vea &=& - \frac{m_{11}}{m_{12}} \uea  \nn \\
\veb &=& \frac{\drel}{n_{22}} - \frac{n_{21}}{n_{22}} \ueb\,. \nn
\eeq
Equivalent results hold when $\lambda = 0$ if the determinant is set equal
to zero. Eliminating the $v$'s in terms of the $u$'s yields
\beq
\left[ u_{1,\skk} (x) v_{1,0} (x)^{*} - u_{1,0} (x)^{*}
v_{1,\skk} (x) \right]^{1}_{0} & = & - \ \frac{u_{1,0} (1)^{*}}{n_{22}}\,\drel
\nn \\
+ \ u_{1,0} (1)^{*} \ueb \left[ \frac{n_{21}}{n_{22}} -
\frac{n_{21}^{*}}{n_{22}^{*}} \right]
& - & u_{1,0} (0)^{*} \uea \left[ \frac{m_{11}}{m_{12}} -
\frac{m_{11}^{*}}{m_{12}^{*}} \right]\,. \nn
\eeq
If we first of all assume that all of the boundary conditions are satisfied,
then the determinant is not present and we see that the general condition
for the operator to be self adjoint is that the ratios $m_{11}/m_{12}$
and $n_{21}/n_{22}$ be real. Since we always have the freedom to multiply
the first line of (\ref{gbc}) by an arbitrary complex number and the
second line by another arbitrary complex number, we can always choose
$m_{12}$ and $n_{22}$ to be real, in which case we deduce that $m_{11}$
and $n_{21}$ should also be real. Therefore if the operator is self-adjoint,
the matrices $M$ and $N$ can always be chosen to be real.

If $m_{12} = 0, n_{22} \neq 0$, then $\uea = 0$ and $u_{1,0} (0) =
0$. The above expression then tells us only that $n_{21}/n_{22}$
must be real if the operator is to be self-adjoint. But now
$m_{11}$ and $n_{22}$ may be chosen to be real, and we once again
find that $M$ and $N$ may be taken to be real. The remaining cases
where $n_{22} = 0$ may be treated in the same way.

In summary, when $\det M = \det N = 0$, if the operator is self-adjoint then
$M$ and $N$ may always be chosen to be real, and
\beq
\left[ u_{1,\skk} (x) v_{1,0} (x)^{*} - u_{1,0} (x)^{*}
v_{1,\skk} (x) \right]^{1}_{0} = \nn
\eeq
\beq
\left\{ \begin{array}{ll}
- \frac{u_{1,0} (1)^{*}}{n_{22}}\,\drel , & \mbox{\ if $n_{22} \neq 0$} \\ \\
+ \frac{v_{1,0} (1)^{*}}{n_{21}}\,\drel , & \mbox{\ if $n_{21} \neq 0$\, .}
\end{array} \right.
\label{appA_i} \eeq Through the boundary condition for the zero
mode, these two forms are clearly equivalent if both $n_{21}$ and
$n_{22}$ are non-zero.

\medskip

\item[(ii)] Suppose that $\det N \neq 0$. Then multiplying
(\ref{modified_gbcs}) by $N^{-1}$ and taking $j=1$ gives
\beq
{ \ueb \choose \veb } &=& \frac{1}{\det N}
\left(
\begin{array}{cc}
n_{22} & - n_{12} \\
- n_{21} & n_{11}
\end{array}
\right)
{ 0 \choose \drel }
\nn \\ \nn \\
&-&
\left(
\begin{array}{cc}
d_{11} & d_{12} \\
d_{21} & d_{22}
\end{array}
\right)
{ \uea \choose \vea }\,,
\nn
\eeq
where the $d_{ij}$ are the elements of the matrix $D \equiv N^{-1} M$.
Substituting for $\ueb$ and $\veb$ (and their $\lambda = 0$
counterparts, which do not contain the $\drel$ term) gives
\beq
\left[ u_{1,\skk} (x) v_{1,0} (x)^{*} - u_{1,0} (x)^{*}
v_{1,\skk} (x) \right]^{1}_{0} \nn
\eeq
\beq
&=& - \frac{\left( n_{11} u_{1,0} (1)^{*}
+ n_{12} v_{1,0} (1)^{*} \right)}{\det N}\,\drel \nn \\ \nn \\
&+& u_{1,0} (0)^{*} \vea \left\{ 1 + d_{12} d_{21}^{*} - d_{11}^{*}
d_{22} \right\} + v_{1,0} (0)^{*} \vea \left\{ d_{12} d_{22}^{*} - d_{12}^{*}
d_{22} \right\} \nn \\ \nn \\
&-& v_{1,0} (0)^{*} \uea \left\{ 1 + d_{12}^{*} d_{21} - d_{11}
d_{22}^{*} \right\} - u_{1,0} (0)^{*} \uea \left\{ d_{21} d_{11}^{*}
- d_{21}^{*} d_{11} \right\}\,. \nn
\eeq
If all boundary conditions were satisfied, the first term on the right-hand
side of this expression would be absent, and the general conditions for
the operator to be self-adjoint are given by the vanishing of the brackets
in the four remaining terms:
\beq
d_{11}^{*} d_{22} - d_{12} d_{21}^{*} &=& 1 \ \ ;\ \
d_{11} d_{21}^{*} = d_{11}^{*} d_{21} \nn \\
d_{11} d_{22}^{*} - d_{12}^{*} d_{21} &=& 1 \ \ ;\ \ d_{12}
d_{22}^{*} = d_{12}^{*} d_{22}\,. \label{conditions_1} \eeq
Examination of the conditions (\ref{conditions_1}) shows that they
may be written in the alternative form \beq d_{ij} = r_{ij} e^{i
\alpha}, \ \ r_{11} r_{22} - r_{12} r_{21} = 1, \ \ 0 \leq \alpha
< 2\pi, \ \ r_{ij} \in \reals . \label{conditions_2} \eeq In other
words, $D=R e^{i\alpha}$, where $R$ is a $2 \times 2$ real matrix
with entries $r_{ij}$ and $\det R = 1$. Therefore if we multiply
(\ref{gbc}) by $N^{-1}$ we see that we may take $M = R
e^{i\alpha}, N = I_{2}$, or equivalently if we multiply by a real
non-singular matrix $N_{R}$, $M = M_{R} e^{i\alpha}, N = N_{R}$
where $M_{R}= N_{R} R$. Note that $\det M_{R} = \det N_{R}$.
\end{itemize}

\bigskip

\section{Asymptotic behaviour of solutions at endpoints}
\label{appB}
\setcounter{equation}{0}

In this appendix we are going to analyse the behaviour of
$(d/d\lambda) \ln\det [M + NE_{j,\slambda} (1)]$ for $\Im\slambda
\to \pm \infty$ as it is needed in equation (\ref{zeta_diff}). We
will omit the index $j$ and consider the general Sturm-Liouville
problem \beq L = -\frac d {dx} \left( P(x) \frac d{dx} \right) +
R(x) , \nn \eeq with the boundary conditions (\ref{gbc}) imposed.
The zeta function associated with this problem is then given by
equation (\ref{zeta_def}), \beq \zeta _L (s) = \frac 1 {2\pi i}
\int\limits_\gamma d\lambda \,\, \lambda^{-s} \frac d {d\lambda}
\ln \det [ M + N E_{\slambda} (1)]. \nn \eeq The meromorphic
structure of the zeta function is determined by the
large-$\Im\slambda $ behaviour of the integrand. The results in
\cite{olve97b} suggest that as $\Im \slambda\to\pm\infty$ the
asymptotic expansion has the general form \beq \frac d {d\lambda}
\ln \det [ M + N E_{\slambda} (1) ] \sim \sum_{n=1} ^\infty
(\slambda) ^{-n} A_{n-1}. \label{ap1} \eeq Note that exponentially
small terms have been dropped.

We will now show that the coefficients $A_{n-1}$ are related to the
associated heat kernel coefficients and use this correspondence to prove that
the first two coefficients do not depend on $R(x)$. As a consequence we can
trivially conclude the behaviour (\ref{asympquot}).

We start summarising some well known facts about the heat kernel
coefficients and their relationship to the zeta function
\cite{gilk95b,kir01}. The heat trace is defined as \beq K(t) =
\sum_l e^{-\lambda_l t} ,\nn \eeq where $\lambda_{l}$ are the
eigenvalues of the operator under consideration. As $t\to 0$, this
sum clearly diverges, since we are summing over infinitely many
eigenvalues. The behaviour as $t\to 0$ may be extracted from a
classical theorem of Weyl \cite{weyl12-71-441}, which, in the
present context, states that for a second order elliptic
differential operator the eigenvalues behave asymptotically for
$l\to \infty$ as \beq \lambda_l^{1/2} \sim \frac{\pi l}{\int_0^1
dx \frac 1 {\sqrt{P(x)}} }. \nn \eeq With the help of a
resummation, \beq \sum_{l=-\infty} ^\infty e^{-tl^2} = \sqrt{\frac
\pi t} \sum_{l=-\infty} ^\infty e^{-\frac{\pi^2 l^2}t}, \nn \eeq
it is seen that this implies $K(t) = {\cal O}(t^{-1/2})$. In more
detail one can show the asymptotic $t\to 0$ behaviour \beq K(t)
\sim \sum_{j=0} ^\infty a_j t^{(j-1)/2}, \label{ap2} \eeq where
exponentially small terms as $t\to 0$ have been neglected. Here,
$a_l$ are the so-called heat kernel coefficients. They depend on
$P(x)$, $R(x)$, and on the boundary conditions imposed. We have,
for example, \beq a_0 = (4\pi)^{-1/2} \int_0^1 dx \,\, \frac 1
{\sqrt{P (x)}} , \quad a_1 = c(M,N) , \nn \eeq where the constant
$c(M,N)$, as indicated, depends on the boundary condition imposed.
The next coefficient $a_2$ involves the dependence on $R(x)$. As
this is of no relevance for us, we do not display higher
coefficients.

The heat kernel coefficients determine the residues and certain
function values of the zeta function. To show how the relationship
is derived we assume that no zero modes are present; otherwise, in
the following calculations we have to exclude them explicitly.

First by definition
\beq
\zeta _L (s) = \sum_{l=0}^\infty \lambda_l^{-s} = \frac 1 {\Gamma (s)}
\sum_{l=0} ^\infty \int_0^\infty dt \,\, t^{s-1} e^{-\lambda_l t} =
\frac 1 {\Gamma(s)} \int_0^\infty dt \,\, t^{s-1} K(t) ,
\nn
\eeq
valid for $\Re s > 1/2$. As is clear, the meromorphic structure of
$\zeta _L (s)$ is related to the $t\to 0$ behaviour of $K(t)$. Thus the
poles of $\zeta _L (s) \Gamma (s)$ are determined by the integrals
\beq
\int_0^1 dt \,\, t^{s-1} \sum_{j=0} ^\infty a_j
t^{(j-1)/2}.
\nn
\eeq
In detail we have
\beq \mbox{Res } \zeta _L(z) &=& \frac{a_{1-2z}} {\Gamma (z)}
\quad \mbox{for } z=\frac 1 2, -\frac{2l+1} 2 , \,\,\, l\in \nats_0,
\nn
\\
\zeta _L (-q) &=& (-1) ^q q! a_{1+2q} \quad \mbox{for }q\in \nats_0 .
\label{zetheat}
\eeq
The asymptotic expansion (\ref{ap1}) determines the above properties of the
zeta function and thus relates $A_n$ and $a_n$. Proceeding as before, see
(\ref{zetcont}), we shrink the contour to the branch cut at the
angle $\theta$. For the case without zero modes, as $\lambda  \to
0$ we have the behaviour $\lambda ^{-s}$ and as $\lambda  \to
\infty$ we have $\lambda  ^{-s-1/2}$. The $\lambda \to 0$
behaviour imposes $\Re s <1$, whereas the $\lambda  \to \infty$
behaviour imposes $\Re s > 1/2$. This shows, that the
representation, as given, is valid for $1/2 <\Re s <1$. It also
shows, that the residues and function values, (\ref{zetheat}),
which all lie to the left of $\Re s >1/2$, are solely determined by the
large-$\lambda $ behaviour. Keeping only the relevant terms to reproduce
(\ref{zetheat}), we continue
\beq \zeta _L (s) &\sim & e^{is (\pi -\theta )} \frac{
\sin (\pi s)} \pi \int_1^\infty d\lambda \,\, \lambda ^{-s}
e^{i\theta} \sum_{n=1} ^\infty e^{-in\theta /2}
\lambda ^{-n/2} A_{n-1}
\nn\\
&=&  e^{is (\pi -\theta )} \frac{ \sin (\pi s)} \pi \sum_{n=1}
^\infty e^{i\theta (1-n/2)} \frac{ A_{n-1}} {s-1+n/2},
\label{strucana} \eeq which can be analysed easily in the whole
complex plane. We see that for $n$ odd, \beq \mbox{Res } \zeta _L
\left( 1-\frac n 2 \right) = \frac i \pi A_{n-1} ,\nn\eeq which
shows \beq A_{n-1} = -i\pi \frac{ a_{n-1}} {\Gamma (1-n/2)},
\label{asym1} \eeq whereas for $n$ even, \beq \zeta _L \left(
1-\frac n 2\right) = A_{n-1} , \nn \eeq and so \beq A_{n-1} = (-1)
^{n/2-1}(n/2-1)! a_{n-1} . \label{asym2} \eeq In particular, \beq
A_0 = -\frac{i\pi a_0} {\Gamma (1/2)} = -\frac i 2 \int_0^1 dx
\,\, \frac 1 {\sqrt{ P(x)}} , \quad A_1 = a_1 = c(M,N), \nn \eeq
and (\ref{asympquot}) follows.

Note, that in (\ref{strucana}) we have $e^{i\theta/2}\slambda$
with positive imaginary part. A negative imaginary part, such as
in $-e^{i\theta/2}\slambda$, changes the sign of $A_{n-1}$ for $n$
odd.

\begin{figure}[ht]
\setlength{\unitlength}{1cm}

\begin{center}

\begin{picture}(20,10)(0,0)

\put(0,0){\setlength{\unitlength}{1.0cm}
\begin{picture}(10,7.5)
\thicklines

\put(0,4){\vector(1,0){10}} \put(5.0,0){\vector(0,1){8}}
\put(5.0,4.0){\line(-1,2){2.0}} \put(.7,8){{\bf cut for
$\lambda^{-s}$}} \multiput(6.5,4)(.4,0){8}{\circle*{.15}}
\multiput(2.5,4)(.4,0){3}{\circle*{.15}}
\put(8.0,7.5){{\bf $\kk$-plane}} \put(4.75,4.5){\line(-1,2){1.75}}
\put(4.75,4.55){\line(-1,2){1.7}} \put(5.0,4.0){\circle{1.10}}
\put(5.55,3.75){\vector(-1,-1){.3}} \put(5.55,3.0){${\bf
\gamma_1}$} \put(4.0,5.0){${\bf \gamma_2}$}

\end{picture}}

\end{picture}

\caption{\label{fig2}Contour $\gamma$ in the complex plane.}

\end{center}

\end{figure}
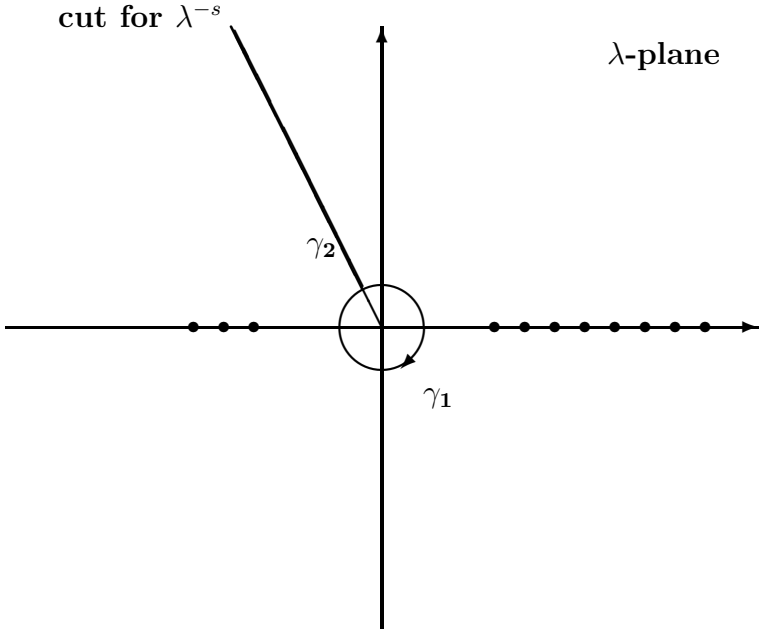

If there are zero modes, say $r$ in number, the equation for $a_1$
changes slightly. First, given we exclude the zero mode from the
definition of the zeta function, we have now \beq \zeta _L (s) =
\frac 1 {\Gamma (s)} \int\limits_0^\infty dt \,\, t^{s-1} (K(t) -
r) , \quad \Re s > \frac 1 2 .\nn\eeq This shows, that
(\ref{zetheat}) remains unchanged apart from \beq \zeta _L (0) =
a_1 -r .\nn\eeq Given there are $r$ zero modes, for $|\lambda |
\ll 1$, we have \beq \frac d {d\lambda} \ln \det [ M + N
E_{\slambda} (1) ] \sim \frac r \lambda + ... \nn\eeq Repeating
the discussion below (\ref{zetheat}), we see that this time the
$\lambda \to 0$ behaviour imposes $\Re s < 0$, which contradicts
the condition $\Re s >1/2$ from $|\lambda | \to \infty$. Therefor,
we cannot shrink the contour to the cut, but instead use the
contour given in Figure \ref{fig2}, consisting of a small circle
$\gamma_1$ of radius $\epsilon$, and of $\gamma_2$ being the part
of $\gamma$ shrunk to the cut.

Along the contour $\gamma_2$ we can proceed as previously and
obtain \beq \zeta _{L,\gamma_2} (s) = e^{is(\pi -\theta)}
\frac{\sin (\pi s)} \pi \int\limits_\epsilon ^\infty d\lambda \,\,
\lambda ^{-s} \frac d {d\lambda} \ln \det [M + N E_{e^{i\theta /2}
\slambda} (1) ] . \label{zero1} \eeq For the contributions along
the circle $\gamma_1$ we obtain \beq \zeta _{L, \gamma_1} (s) = -
e^{is (\pi -\theta )} \frac{\sin (\pi s)} \pi \frac{r\epsilon
^{-s}} s .\label{zero2}\eeq The contribution of (\ref{zero1}) to
the quantities in (\ref{zetheat}) are evaluated precisely as
before. In addition, as $s\to 0$, (\ref{zero2}) produces \beq
\zeta_{L,\gamma_1} (0) = -r.\nn\eeq As a result, the asymptotic
behaviour is determined again through equations (\ref{asym1}) and
(\ref{asym2}).

\bigskip

\section{Zero modes in systems of differential operators}
\label{appC}
\setcounter{equation}{0}

In Section \ref{systems} we introduced the vectors $b$ and $b_c$
and the matrices ${\cal Z}$ and ${\cal Z} _c$. The $2r$
dimensional vectors $b$ and $b_c$ together contain the $4r$
boundary data: $r$ data coming from each of $u_{1,\sla} (0)$,
$v_{1,\sla} (0)$, $u_{1,\sla} (1)$ and $v_{1,\sla} (1)$. The $2r
\times 2r$ matrices ${\cal Z}$ and ${\cal Z} _c$ together contain
all the elements of the matrices $M$ and $N$ but rearranged in a
way which corresponds to the organisation of the boundary data in
$b$ and $b_c$. The purpose of this appendix is to make explicit
the notation required to describe which of the $2r$ boundary data
goes into $b$ and which goes into $b_c$ and which of the elements
of $M$ and $N$ go into ${\cal Z}$ and which go into ${\cal Z} _c$.

To do this we introduce indices $i,j,k,l$ and permutations $\{a_1,...,a_r\}$,
$\{b_1,...,b_r\}$, $\{c_1,...,c_r\}$ and $\{d_1,...,d_r\}$ as in Section
\ref{systems}. Let us recall that these index groups are such that $m_{a_i}$
acts on boundary data in $u_{1,\sla} (0)$, $m_{b_j}$ acts in $v_{1,\sla} (0)$,
$n_{c_k}$ acts in $u_{1,\sla} (1)$, and $n_{d_l}$ acts in $v_{1,\sla} (1)$.
So if
\beq
b=\left( \begin{array}{c} u_{1,\sla , a_1} (0) \\
... \\
u_{1,\sla , a_i } (0) \\
v_{1,\sla , b_1-r} (0) \\
...\\
v_{1,\sla , b_j -r} (0) \\
u_{1,\sla , c_1 } (1) \\
...\\
u_{1,\sla , c_k } (1) \\
v_{1,\sla , d_1 -r } (1) \\
...\\
v_{1,\sla , d_l -r} (1) \end{array} \right), \quad b_c =
\left( \begin{array}{c} u_{1,\sla , a_{i+1}} (0) \\
... \\
u_{1,\sla , a_r } (0) \\
v_{1,\sla , b_{j+1}-r} (0) \\
...\\
v_{1,\sla , b_r -r} (0) \\
u_{1,\sla , c_{k+1} } (1) \\
...\\
u_{1,\sla , c_r } (1) \\
v_{1,\sla , d_{l+1}-r } (1) \\
...\\
v_{1,\sla , d_r-r } (1) \end{array} \right),
\nn
\eeq
then
\beq
{\cal Z} =\left(
\begin{array}{cccccc} m_{1a_1} & ... & m_{1a_i} & m_{1b_1} &
... & m_{1b_j} \\
... & ...& ...& ...& ...& ... \\
m_{(2r)a_1} & ... & m_{(2r)a_i} & m_{(2r)b_1} & ... & m_{(2r)b_j}
\end{array}
\right. \\ \nn \\ \nn
\left.
\begin{array}{cccccc} n_{1 c_1} & ... & n_{1 c_k} & n_{1 d_1} & ...
& n_{1 d_l} \\
...& ...& ...& ...& ...& ... \\
n_{(2r) c_1} & ... & n_{(2r) c_k} & n_{(2r) d_1} & ... & n_{(2r)d_l}
\end{array}
\right),
\eeq
and
\beq
{\cal Z}_c =\left(
\begin{array}{cccccc} m_{1a_{i+1}} & ... & m_{1a_r} & m_{1b_{j+1}} &
... & m_{1b_{r}} \\
... & ...& ...& ...& ...& ... \\
m_{(2r)a_{i+1}} & ... & m_{(2r)a_r} & m_{(2r)b_{j+1}} & ... &
m_{(2r)b_{r}}
\end{array}
\right. \\ \nn \\ \nn
\left.
\begin{array}{cccccc} n_{1 c_{k+1}} & ... & n_{1 c_r}
& n_{1 d_{l+1}} & ... & n_{1 d_{r}} \\
...& ...& ...& ...& ...& ... \\
n_{(2r) c_{k+1}} & ... & n_{(2r) c_r} & n_{(2r)
d_{l+1}} & ... & n_{(2r) d_r}
\end{array}
\right) \eeq The notation is such that if one of the indices
$i,j,k,l$ equals zero, then the corresponding entries above are
simply absent. It is clear from (\ref{datasplit}) that $b$ can be
expressed through $b_c$ only if the matrix ${\cal Z}$ is
invertible. If this is the case then, for a particular choice of
${\cal Z}$, (\ref{datasplit}) allows us to express the $2r$ data
$b$ in terms of the complementary $2r$ data $b_c$ as \beq
\left( \begin{array}{c} u_{1,\sla , a_1} (0) \\
... \\
u_{1,\sla , a_i } (0) \\
v_{1,\sla , b_1-r} (0) \\
...\\
v_{1,\sla , b_j -r} (0) \\
u_{1,\sla , c_1 } (1) \\
...\\
u_{1,\sla , c_k } (1) \\
v_{1,\sla , d_1 -r} (1) \\
...\\
v_{1,\sla , d_l -r} (1) \end{array} \right) = - {\cal Z} ^{-1}
{\cal Z}_c
\left( \begin{array}{c} u_{1,\sla , a_{i+1}} (0) \\
... \\
u_{1,\sla , a_r } (0) \\
v_{1,\sla , b_{j+1}-r} (0) \\
...\\
v_{1,\sla , b_r -r} (0) \\
u_{1,\sla , c_{k+1} } (1) \\
...\\
u_{1,\sla , c_r } (1) \\
v_{1,\sla , d_{l+1}-r } (1) \\
...\\
v_{1,\sla , d_r-r } (1) \end{array} \right) + {\cal Z} ^{-1}
\left(
\begin{array}{c} 0
\\ ...\\...\\...\\...\\...\\...\\...\\...\\...\\0\\\det ( M+N
E_{1,\sla} (1) ) \end{array} \right).
\label{4}
\eeq
We may now substitute the $2r$ quantities on the left-hand side of (\ref{4})
into the left-hand side of (\ref{starting_point}). As we have argued for the
case $r=1$, all terms that do not depend explicitly on the term containing
$\det (M+N E_{1,\sla} (1))$ have to cancel each other due to the
self-adjointness of the boundary value problem. Without attempting to state
the conditions required for $M$ and $N$ (this does not do any harm simply
because they are not needed), we keep only terms that do depend on the term
$\det (M+N E_{1,\sla} (1))$, knowing the others {\it have} to cancel. In this
way we arrive at
\beq
[u_{1,\sla} (x) v_{1,0} (x) ^* - u_{1,0}
(x)^* v_{1,\sla} (x) ]^1_0 &=& \det (M+N E_{1,\sla} (1))\times \nn\\
& &\hspace{-3cm}\left\{\sum_{\alpha =1} ^k {\cal
Z}_{(i+j+\alpha)(2r)} ^{-1} v_{1,0,c_\alpha} (1) ^* - \sum_{\alpha
=1} ^l{\cal
Z}_{(i+j+k+\alpha)(2r)} ^{-1} u_{1,0,d_\alpha -r} (1) ^* \right.\nn\\
& &\hspace{-3cm}\left. -\sum_{\alpha =1} ^i {\cal Z}_{\alpha (2r)}
^{-1} v_{1,0,a_\alpha} (0) ^* + \sum_{\alpha =1} ^j {\cal
Z}_{(i+\alpha) (2r)} ^{-1} u_{1,0,b_\alpha -r} (0)
^*\right\}\,,
\label{app_5}
\eeq
where ${\cal Z}_{\beta \gamma} ^{-1}$ refers to the $(\beta \gamma)$-component
of ${\cal Z}^{-1}$. This is of the desired form (\ref{bit_of_three}) with
${\cal B}$ given by (\ref{Bminus1}).

\end{appendix}

\end{document}